\documentclass[11pt]{llncs}
\usepackage[utf8]{inputenc}
\usepackage[margin=1in]{geometry}

\usepackage{amsfonts,amsmath,amssymb}
\usepackage{graphicx}
\usepackage{mdframed}
\usepackage{tikz}
\usepackage[ruled]{algorithm2e}
\usepackage[colorinlistoftodos,textsize=small]{todonotes}
\usetikzlibrary{matrix}
\usetikzlibrary{calc}
\usetikzlibrary{shapes}

\spnewtheorem{thm}{Theorem}{\bfseries }{\itshape }
\spnewtheorem{lem}[thm]{Lemma}{\bfseries }{\itshape }
\spnewtheorem{cor}[thm]{Corollary}{\bfseries }{\itshape }
\spnewtheorem{clm}{Claim}{\bfseries }{\itshape }
\numberwithin{clm}{thm}
\spnewtheorem{defn}{Definition}{\bfseries }{\itshape }

\newcommand{\Fig}[1]{pics/#1}

\newcommand{\ShowTODO}[1]{{#1}}
\renewcommand{\ShowTODO}[1]{}

\newenvironment{dproof}{\begin{proof}}{\qed\end{proof}}

\newcommand{\E}[3]{E_{#1}(#2,#3)}
\newcommand{\Model}{\mathcal{M}}
\newcommand{\CompSec}[1]{\mathcal{S}_{#1}}
\newcommand{\Def}[1]{{\sf#1}}
\newcommand{\Op}{\text{{\tt(}}}
\newcommand{\Cp}{\text{{\tt)}}}
\newcommand{\ub}{\text{{\tt.}}}
\newcommand{\Ab}{{\sf A}}
\newcommand{\Cb}{{\sf C}}
\newcommand{\Gb}{{\sf G}}
\newcommand{\Ub}{{\sf U}}
\newcommand{\wP}{u}
\newcommand{\wS}{v}
\newcommand{\Sp}{S_{u}}
\newcommand{\Ss}{S_{v}}

\newcommand{\True}{{\sf True}}
\newcommand{\False}{{\sf False}}

\DeclareMathOperator*{\argmin}{argmin}

\newcommand{\Result}[1]{{\bf R#1}}

\newcommand{\MotifA}{{m}_{5}}
\newcommand{\MotifB}{{m}_{3\,\circ}}

\newcommand{\RNAFold}{\ensuremath{\text{\sf RNA-FOLD}}}
\newcommand{\UniqueFold}{\ensuremath{\text{\sf UNIQUE-FOLD}}}
\newcommand{\RNADesign}{\ensuremath{\text{\sf RNA-DESIGN}}}

\newcommand{\Designable}[1]{\text{\sf Designable}(#1)}
\newcommand{\Paired }[1]{\text{\sf Paired}(#1)}
\newcommand{\Unpaired }[1]{U_{#1}}
\newcommand{\MaxDeg}[1]{D({#1})}
\newcommand{\Sep}{{\sf Sep}}

\newcommand{\LevelFun}{L}
\newcommand{\Level}[1]{\LevelFun(#1)}

\newcommand{\Prefix}[2]{#2_{[1,#1]}}

\newcommand{\Substring}[3]{#3_{[#1,#2]}}

\tikzstyle{bp}=[thick,draw=blue,in =95,out=85,looseness=1.3]
\tikzstyle{altbp}=[thick,draw=red,dashed,in =-95,out=-85,looseness=.5]
\tikzstyle{cell}=[inner sep=2]

\def\mySecStr#1{\expandafter {\tt #1}\& }
\def\mySecStrAll#1{\ifx#1\mySecStrAll\else\mySecStr#1\expandafter\mySecStrAll\fi}
\def\mySeq#1{\expandafter {\relsize{-2}\sf #1}\&}
\def\mySeqAll#1{\ifx#1\mySeqAll\else\mySeq#1\expandafter\mySeqAll\fi}

\newcommand{\Problem}[3]{\begin{mdframed}[backgroundcolor=lightgray!10,
  linecolor=black,
  linewidth=0pt]
\noindent#1 problem\\
{\sf Input:} #2\\
{\sf Output:} #3
\end{mdframed} 
}
\newcommand{\WM}{\ensuremath{{\cal W}}}

\newcommand{\RNA}[4][]{
\begin{tikzpicture}[baseline={([yshift=-.5ex]rna)}]
  \matrix[matrix of nodes,nodes=cell,ampersand replacement=\&] (rna){
		\mySecStrAll #2 \mySecStrAll\\
		\ifthenelse{\equal{#3}{}}{}{ \mySeqAll #3 \mySeqAll\\}
		};
\ifthenelse{\equal{#4}{}}{}{%
	\foreach \x/\y in {#4}{\draw (rna-1-\x) edge[bp] (rna-1-\y);}%
}
\ifthenelse{\equal{#1}{}}{}{%
	\foreach \x/\y in {#1}{\draw (rna-2-\x) edge[altbp] (rna-2-\y);}%
}
\end{tikzpicture}}

\tikzstyle{jbp0}=[thick,draw=blue,in =90,out=90,looseness=1]
\tikzstyle{jbp1}=[thick,draw=red,in =-90,out=-90,looseness=1]
\tikzstyle{jbp2}=[thick,draw=red,dashed,in =-90,out=-90,looseness=1]
\tikzstyle{jbp3}=[thick,draw=blue,dashed,in =-90,out=-90,looseness=1]
\newcommand{\JRNA}[6][]{
\begin{tikzpicture}[baseline={([yshift=-.5ex]rna)}]
  \matrix[matrix of nodes,nodes=cell,ampersand replacement=\&] (rna){
		\ifthenelse{\equal{#2}{}}{}{ \mySeqAll #2 \mySeqAll\\}
		};
\ifthenelse{\equal{#3}{}}{}{%
	\foreach \x/\y in {#3}{\draw (rna-1-\x) edge[jbp0] (rna-1-\y);}%
}
\ifthenelse{\equal{#4}{}}{}{%
	\foreach \x/\y in {#4}{\draw (rna-1-\x) edge[jbp1] (rna-1-\y);}%
}
\ifthenelse{\equal{#5}{}}{}{%
	\foreach \x/\y in {#5}{\draw (rna-1-\x) edge[jbp2] (rna-1-\y);}%
}
\ifthenelse{\equal{#6}{}}{}{%
	\foreach \x/\y in {#6}{\draw (rna-1-\x) edge[jbp3] (rna-1-\y);}%
}
\ifthenelse{\equal{#1}{}}{}{%
	\foreach \x/\y/\b/\a in {#1}{%
		\coordinate (X) at ($ (rna-1-\x)!.5!(rna-1-\y) $);%
		\draw ($ (X) - (0,\b) $) -- ($ (X) + (0,\a) $);%
	}%
}
\end{tikzpicture}}

\begin{document}

\title{Combinatorial RNA Design:\\Designability and Structure-Approximating Algorithm}

\author{Jozef Hale\v{s}\inst{1}\and J\'an Ma\v{n}uch\inst{1,3}\and Yann Ponty\inst{1,2} \and Ladislav Stacho\inst{1}}

\institute{
Department of Mathematics, Simon Fraser University, Canada
\and
Pacific Institute for Mathematical Sciences, CNRS UMI3069, Canada
\and 
Department of Computer Science, University of British Columbia, Canada
}

\maketitle{}

\begin{abstract} 
In this work, we consider the {\em Combinatorial RNA Design problem}, a {\em minimal} instance of the RNA design problem which aims at finding a sequence that admits a given target as its unique base pair maximizing structure.
  We provide complete characterizations for the structures that can be designed using restricted alphabets. 

Under a classic four-letter alphabet, we provide a complete characterization of designable structures without unpaired bases. When unpaired bases are allowed, we provide partial characterizations for classes of designable/undesignable structures, and show that the class of designable structures is closed under the stutter operation. Membership of a given structure to any of the classes can be tested in linear time and, for positive instances, a solution sequence can also be generated in linear time.

Finally, we consider a structure-approximating version of the problem that allows to extend bands (helices) and, assuming that the input structure avoids two motifs, we provide a linear-time algorithm that produces a designable structure with at most twice more base pairs than the input structure. 
\end{abstract}

\section{Introduction}
\label{sec:introduction}
RiboNucleic Acids (RNAs) are biomolecules which act in almost every aspect of cellular life, and can be abstracted as a sequence of nucleotides, i.e., a string over the alphabet $\{\Ab,\Ub,\Cb,\Gb\}$.
Due to their versatility, and the specificity of their interactions, they are increasingly being used as therapeutic agents~\cite{Wu2014}, and as building blocks for the emerging field of synthetic biology~\cite{Rodrigo2013,Takahashi2013}. A substantial proportion of the functional roles played by RNA rely on interactions with other molecules to activate/repress dynamical properties of some biological system, and ultimately require the adoption of a specific conformation. Accordingly, RNA bioinformatics has dedicated much effort to developing energy models~\cite{Mathews1999,Turner2010} and algorithms~\cite{Nussinov1980,Zuker1981} to predict the \Def{secondary structure of RNA}, a combinatorial description of the conformation adopted by an RNA which only retains interacting positions, or base pairs.
Historically, structure prediction has been addressed as an optimization problem, whose expected output is a secondary structure which minimizes some notion of free-energy~\cite{Nussinov1980,Zuker1981}. The performances of the RNA folding prediction problem have now reached a point where {\em in silico} predictions have become generally reliable~\cite{Mathews1999}, allowing for large scale studies and fueling the discovery of an increasing number of functional families~\cite{Griffiths-Jones2003}.

Due to the existence of expressive, yet tractable, energy models, coupled with promising applications in multiple fields (pharmaceutical research, natural computing, biochemistry\ldots), a wide array of computational methods~\cite{Hofacker1994,Busch2006,Aguirre-Hernandez2007,Dai2009,Avihoo2011,Taneda2011,Zadeh2011,Levin2012,Lyngso2012,Garcia-Martin2013a,HoenerZuSiederdissen2013,Reinharz2013,Zhou2013,Esmaili-Taheri2014}
 have been proposed to tackle the natural inverse version of the structure prediction, the \Def{RNA design problem}. In this problem, one attempts to perform the {\em in silico} synthesis of artificial RNA sequences,  performing a predefined biological function {\em in vitro} or {\em in vivo}. Given the prevalence of structure in the function of an RNA, one of the foremost goal of RNA design (sometimes named  \Def{inverse folding} in the literature) is that the designed sequence should fold into a predefined secondary structure. In other words, it should not be challenged by alternative stable structures having similar or lower free-energy. 

Despite a rich, fast-growing, body of literature dedicated to the problem, there is currently no exact polynomial-time algorithm for the problem. Moreover, the complexity of the problem remains open (see Section~\ref{sec:conclusion} for details). It can be argued that this situation, quite exceptional in the field of computational biology, partly stems from the intricacies of the Turner free-energy model~\cite{Turner2010} which associates experimentally-determined energy contributions to $\sim2.4\times 10^4$ structure/sequence motifs. This motivates a reductionist approach, where one studies an idealized version of the RNA design problem, lending itself to algorithmic intuitions, while hopefully retaining the presumed difficulty of the original problem.

In this work, we introduce the {\em Combinatorial RNA Design problem}, a {\em minimal} instance of the RNA design problem which aims at finding a sequence that admits the target structure as its unique base pair maximizing structure. 
After this short introduction, Sec.~\ref{sec:defs} states definitions and problems. In Sec.~\ref{sec:results}, we summarize our results, some of which are proven in Sec.~\ref{sec:proofs}. Finally, we conclude in Sec.~\ref{sec:conclusion} with some remarks, open problems and future extensions of this work.

\section{Definitions and notations}\label{sec:defs}

\begin{figure}[tb]
{\centering
\includegraphics[width=\linewidth]{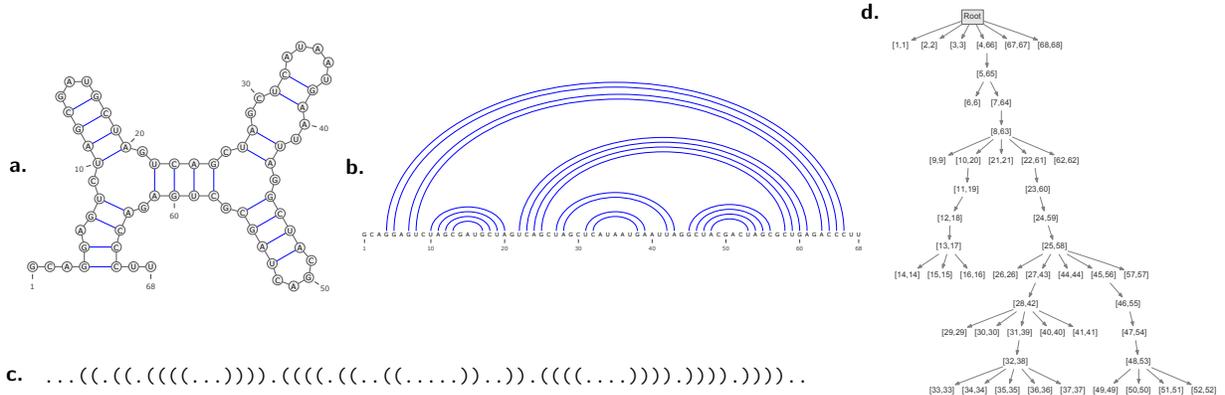}
\\}
\caption{Four equivalent representations for an RNA secondary structure of length $68$, consisting of $20$ base pairs forming $7$ bands: outer-planar graph (a.), arc-annotated representation (b.), parenthesized expression (c.), and tree representation (d.)}
\end{figure}

\paragraph{RNA secondary structure.}
An RNA can be encoded as a \Def{sequence} of nucleotides, i.e., a string $w = w_{1}\cdots w_{|W|}\in\{\Ab,\Ub,\Cb,\Gb\}^\star$. The \Def{prefix of $w$ of length $i$} is denoted as $\Prefix{i}{w}$ and $|w|_{b}$ denotes the number of occurrences of $b$ in $w$.
A \Def{(pseudoknot-free) secondary structure} $S$ on an RNA of length $n$ is a pair $(n,P)$, where $P$ is a set of \Def{base pairs} $\{(l_{i},r_{i})\}_{i=1}^p \subset [1,n]^2$ such that:
\begin{itemize}
\item $\forall i\in [1,p]$, $l_i<r_i$;
\item $\forall i, j\in [1,p], l_{i}\ne l_{j}$, $l_{i}\ne
  r_{j}$, $r_{i}\ne r_{j}$ (each position is involved in at most
  one base pair);
\item $\nexists i,j\in [1,p]$, $l_{i} < l_{j} < r_{i} < r_{j}$ (base pairs $(l_{i},r_{i})$ and $(l_{j},r_{j})$ do not \Def{cross}).
\end{itemize}
The \Def{set of all secondary structures} is denoted by $\CompSec{}$, and its restriction to structures of length $n$ by $\CompSec{n}$. 
The \Def{unpaired positions} $\Unpaired{S}$ in a secondary structure $S = (n,P)$ is the set of indices $k\in[1,n]$ that are not involved in a base pair. A structure $S$ is \Def{saturated} if $\Unpaired{S} = \emptyset $. Given a sequence $w$ and a structure $S = (|w|,P)$, let $u_{i} = \varepsilon $ if $i\in \Unpaired{S}$ and $u_{i} = w_{i}$, otherwise, where $\varepsilon $ is the empty sequence. Define the \Def{$S$-paired restriction of $w$ (paired restriction of $S$)}, denoted as $\Paired{w,S}$ ($\Paired{S}$), as $u_{1}\cdots u_{|w|}$ (respectively, $\{(|u_{1}\cdots u_{i}|,|u_{1}\cdots u_{j}|)\mid (i,j)\in P\}$).
A maximal subset $B = \{(i,j),(i + 1,j - 1),\dots,(i + \ell, j - \ell)\}$ of $P$ for some integer $i,j,\ell $ is called a \Def{band} (sometimes referred to as \Def{helix} or \Def{stem}) of  \Def{size} $\ell = |B|$, of $S = (n,P)$. Note that every base pair belongs to exactly one band. 

\paragraph{Dot-parentheses notation.}
A \Def{well-parenthesized sequence} $s\in \{\Op,\Cp,\ub\}^{*}$ can be used to represent a secondary structure. There is one-to-one correspondence between secondary structures and such well-parenthesized sequences: any base pair $(l,r)\in S$ becomes a pair of corresponding opening and closing parentheses in $s$ at position $l$ and $r$ respectively ($s_l = \Op$ and  $s_r = \Cp$), and any unpaired position $i$ corresponds to a dot ($s_i=\ub$).

\paragraph{$k$-stutter.}
The $k$-\Def{stutter} of a sequence $s$, denoted by $s^{[k]}$ is the result of an independent copy $k$-times of each of the characters in $s$. This operation can be applied to both RNA sequences and structures in the dot-parentheses notation.

\paragraph{Tree representation.}
Alternatively, the \Def{tree representation}, denoted by $T_{S}$, for $S = (n,P)$ is a rooted ordered tree whose vertex set $V_S$ consists of intervals $[l,r]$ for any base pair $(l,r)\in P$, and $[k,k]$ for every $k\in \Unpaired{S}$.
A \Def{virtual root} $[0,n+1]$ is added for convenience. Each $[k,k]$ node is called \Def{unpaired} node, all other nodes (including the root) are called \Def{paired} nodes. The \Def{children} of an interval $I \in V_S$ are the maximal proper subintervals $I'\in V_S$ of $I$ ordered by the left points of the intervals. 
The \Def{degree} of a vertex $I\in V_S$ is the total number of its paired neighbors, including its parent (if any).
We denote by $\MaxDeg{S}$ the maximal degree of nodes in $T_{S}$.

\paragraph{Proper, greedy and separated coloring of the tree representation.}
Consider the tree representation $T_{S}$ of structure $S$. Color every paired node of $T_{S}$ different from the root by black, white or grey color. This coloring is called \Def{proper} if:
\begin{enumerate}
\item 
every node has at most one black, at most one white and at most two grey children;
\item 
a node with color $c$ has at most one child with color $c$;
\item 
a black node does not have a white child and a white node does not have a black child.
\end{enumerate}
A \Def{greedy coloring} of $T_{S}$ is the coloring obtained by recursive application of the following rule starting from the root and continuing towards leaves: if the node is black, color the first paired child black and the remaining paired children grey, if the node is white, color the first paired child white and the remaining paired children grey, otherwise (the grey node or the root), color the first paired child black, second white and the remaining paired children grey. It is easy to check that if the degree of each node is at most four then the greedy coloring is a proper coloring.

Given a proper coloring of $T_{S}$, let the \Def{level} of each node be the number of black nodes minus the number of white nodes on the path from this node to the root. A proper coloring is called \Def{separated} if the two sets of levels, associated with grey and unpaired nodes respectively, do not overlap.

\subsection{Statement of the generic RNA design problem}

Consider an \Def{energy model} $\Model$, which associates a \Def{free-energy} $\E{\Model}{w}{S}\in \mathbb{R}^-\cup \{+\infty\}$ to each secondary structure $S\in \CompSec{|w|}$ for a given RNA sequence $w$.
The \Def{minimum free-energy (MFE) structure prediction} problem is typically defined as follows:
\Problem{\RNAFold$_{\Model}$}%
{RNA sequence $w$}%
{$S_{\Model }^\star (w) := \argmin_{S'\in \CompSec {|w|}} \E{\Model}{w}{S'}\,.$}%

The existence of competing structures, having comparable or lower free-energy for a given RNA, impacts the well-definedness of the folding process. The detection of such situations is therefore of interest, and can be rephrased as follows:
\Problem{\UniqueFold$_{\Model}$}%
{Sequence $w$ + Energy distance $\Delta>0$}%
{\True{} if, for every
$S'\in \CompSec{|w|}\setminus\{S_\Model^\star(w)\}$,
$\E{\Model}{w}{S'} \ge \E{\Model}{w}{S_\Model^\star(w)}+\Delta \,.$\\
\False{} otherwise.}

We can now define the \Def{combinatorial RNA Design problem} as:
\Problem{\RNADesign$_{\Model,\Sigma}$}%
{Secondary structure $S$ + Energy distance $\Delta>0$}%
{RNA sequence $w\in\Sigma^\star$ --- called an \Def{$(\Model ,\Sigma ,\Delta )$-design for $S$} --- such that:\\
$$ \RNAFold_\Model(w)=S\ \text{ and }\ \UniqueFold_\Model(w,\Delta),$$
or {$\varnothing$} if no such sequence exists.}

Structures for which there exists an $(\Model ,\Sigma ,\Delta )$-design are called \Def{$(\Model ,\Sigma ,\Delta )$-designable}. Let $\Designable{\Model ,\Sigma ,\Delta } $ be the set of all such structures. If it is clear from the context, we will usually drop $\Model $, $\Sigma $ and/or $\Delta $.

\begin{figure}[tb]
  \centering
  \begin{minipage}{.55\linewidth}
  \includegraphics[width=\linewidth]{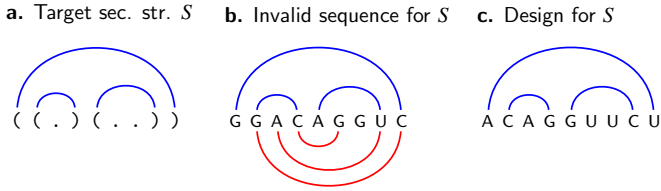}
  \end{minipage}
  \begin{minipage}{.44\linewidth}
  \caption{The combinatorial RNA design problem: Starting from a secondary structure $S$ ({\sf a.}), our goal is to design an RNA sequence which uniquely folds, with maximum number of base pairs, into $S$. The sequence proposed in {\sf b.} is invalid due to the existence of an alternative structure (lower half-plane, red) having the same number of base pairs as $S$. The right-most sequence ({\sf c.}) is a design for $S$.}
  \end{minipage}
\end{figure}

\subsection{Combinatorial design in a simple base pair energy model}

In this work, we adopt a \Def{Watson-Crick} energy model $\WM$, which only allows pairs involving \Def{complementary} letters, i.e., in $\{\Cb,\Gb\}$ and $\{\Ab ,\Ub \}$.

\begin{defn}[Watson-Crick energy model $\WM$]
$$\E{\WM}{w}{S} =
\begin{cases}
  -|S|& \text{if $\forall (l,r)\in S$, $w_{l}$ is complementary with $w_{r}$,}\\
  +\infty& \text{otherwise.}\\
\end{cases}$$
\end{defn}

We say that the structure is \Def{compatible} with a sequence $w$, if $\E{\WM }{w}{S} < +\infty $.

Minimizing $\E{\WM}{w}{S}$ is equivalent to maximizing $|S|$, thus \RNAFold$_{\WM}$ is a classic base pair maximization problem. It can be solved by dynamic programming, historically in $\mathcal{O}(n^3)$ complexity ~\cite{Nussinov1980}, or in $\mathcal{O}(n^3/\log(n))$ current best time complexity~\cite{Frid2010}.
A backtracking procedure reconstructs the MFE structure, and can be easily
adapted to assess the uniqueness of the MFE structure.

\section{Statement of the results}\label{sec:results}

We consider the design problem in a base pairing energy model \WM{} restricted to Watson-Crick base pairs $\{\Cb,\Gb\}$ and $\{\Ab ,\Ub \}$. We set $\Delta=1$, which forbids designed sequence to adopt alternative structures having greater or equal number of base pairs than the target structure.  Let us first characterize the sets $\Designable{\Sigma}$ of designable structures over partial alphabets $\Sigma$. Let $\Sigma_{c,u} $ be an alphabet with $c$ pairs of complementary bases and $u$ bases without a complementary base.

\paragraph{Designability over restricted alphabets.}
\begin{enumerate}
\item[\Result{1}] 
  For every $u\in\mathbb{N}^{+}$, $\Designable{\Sigma_{0,u} } = \{(n,\emptyset)\mid\forall n\in\mathbb{N}\}$;
\item[\Result{2}] 
  $\Designable{\Sigma_{1,0}} = \{S\in\CompSec{} \mid S \text{ is saturated and } \MaxDeg{S}\le 2\} \cup \{(n,\emptyset)\mid\forall n\in\mathbb{N}\}$;
\item[\Result{3}] 
  $\Designable{\Sigma_{1,1}} = \{S\in\CompSec{} \mid \MaxDeg{S}\le 2\}$.
\end{enumerate}
\paragraph{Designability over the complete alphabet $\Sigma_{2,0} =  \{\Ab,\Ub,\Cb,\Gb\}$.}
\begin{enumerate}
\item[\Result{4}] 
  $\Designable{\Sigma_{2,0}} \cap \{S\in\CompSec{} \mid S \text{ is saturated} \} = \{S\in\CompSec{} \mid \MaxDeg{S}\le 4\}\cap \{S\in\CompSec{} \mid S \text{ is saturated} \}$.
\end{enumerate}
When unpaired positions are allowed in the target structure, our characterization is only partial:
\begin{enumerate}
\item[\Result{5}] 
  Let $\MotifA$ represent ``a node having degree more than four'', and $\MotifB$ be ``a node having one or more unpaired children, and degree greater than two'', then 
$$\Designable{\Sigma_{2,0}} \cap  \{S\in\CompSec{} \mid \text{$S$ contains $\MotifA$ or $\MotifB$}\} = \emptyset \,;$$
\item[\Result{6}] 
  Let $\Sep$ be the set of structures for which there exists a separated (proper) coloring of the tree representation, then $\Sep\subset \Designable{\Sigma_{2,0}}$;
\item[\Result{7}] 
  The set of $\Sigma_{2,0}$-designable structures is closed under the $k$-stutter operations:
$$ \forall S\in \CompSec{}, \forall k\in \mathbb{N}^+:\quad S\in \Designable{\Sigma_{2,0}} \implies S^{[k]} \in \Designable{\Sigma_{2,0}}\,.$$
\end{enumerate}

\begin{figure}[tb]
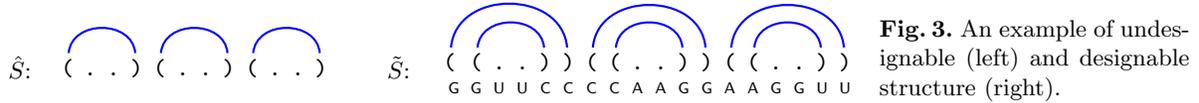

\centering 
\begin{minipage}{0.70\textwidth}
  \centering
  $\hat S$: \RNA[]{(\ub\ub)(\ub\ub)(\ub\ub)}{ }{1/4,5/8,9/12}
  \hspace{0.4cm}
  $\tilde S$: \RNA[]{((\ub\ub))((\ub\ub))((\ub\ub))}{GGUUCCCCAAGGAAGGUU}{1/6,2/5,7/12,8/11,13/18,14/17}
\end{minipage}
\begin{minipage}{0.25\textwidth}
  \caption{An example of undesignable (left) and designable structure (right).}
  \label{fig:struct:input}
\end{minipage}
\vspace*{-3mm}
\end{figure}

\begin{figure}[tb]
{\centering \includegraphics[width=.9\textwidth]{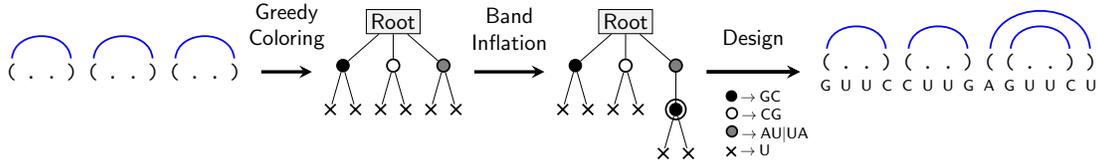}\\}\vspace{-1em}
\caption{Application of the structure-approximating algorithm to the non-designable structure $\hat S$ in Fig.~\ref{fig:struct:input}:
A base pair (circled black node) is inserted in the greedily colored tree, offsetting the levels of white and unpaired nodes (crosses) to even and odd levels respectively, so that the resulting tree is proper/separated, representing a designable structure.}
\label{fig:inflation}
\end{figure}

We note that $S^{[k]}\in \Designable{\Sigma_{2,0}}$ does not imply that $S\in \Designable{\Sigma_{2,0}}$. For instance, it can be verified that $\hat S^{[2]}$ is $\Sigma_{2,0}$-designable, while $\hat S$ is not, cf. Figure~\ref{fig:struct:input}. Membership to the classes described in \Result{1}-\Result{5} can be tested by trivial linear-time algorithms,  which can also be adapted into linear-time algorithms for the \RNADesign$_{\Model,\Sigma}$ problem.
\paragraph{Structure-approximating algorithm.} 
Unfortunately, the absence of $\MotifA$ or $\MotifB$, while necessary, is generally not sufficient to ensure designability. For instance, $\hat S$ in Figure~\ref{fig:struct:input}
clearly does not contain $\MotifA$ or $\MotifB$, yet cannot be designed.
In such cases, the unwanted interactions can be penalized by the duplication of some base pairs. For instance, duplicating the base pairs in the above example yields $\Sigma_{2,0}$-designable structure $\tilde S$.
\begin{enumerate}
\item[\Result{8}] Any structure $S$ without $\MotifA$ and $\MotifB$ can be transformed in $\Theta(n)$ time into a $\Sigma_{2,0}$-designable structure $S'$, by inflating a subset of its base pairs (at most one per band) so that the greedy coloring of the resulting structure is proper and separated, as illustrated by  Figure~\ref{fig:inflation}.
\end{enumerate}
%
%
%
%
%
%
%
\section{Proofs}\label{sec:proofs}
\Result{1} is trivial since, in the absence of complementary letters, the structures without base pairs are the only structures whose energy is not infinite. 
\begin{thm}[Result \Result{4}]
  \label{t:result4}
  A saturated sec. str. $S$ is $\Sigma_{c,0}$-designable if and only if $\MaxDeg{S}\le 2c$.
\end{thm}

\begin{dproof}
First, we will show that the degree condition is necessary. Assume to the contrary that $\MaxDeg{S} > 2c$ and $S$ has a design $w$. Let $[a,b]$ be a vertex with degree $d\ge 2c + 1$ in $T_{S}$. Let $\{[l_{i},r_{i}]\}_{i = 1 }^{d}$ be the (paired) children of $[a,b]$ and the node $[a,b]$ if $[a,b]$ is not the root. Let $L_{i} = l_{i}$ and $R_{i} = r_{i}$ if $[l_{i},r_{i}]$ is a child of $[a,b]$, and $L_{i} = r_{i}$ and $R_{i} = l_{i}$ if it is $[a,b]$. Then among bases $w_{L_{1}},\dots,w_{L_{d}}$ must be a pair of repeated letters. Let $w_{L_{i}} = w_{L_{j}}$ be such a pair with $L_{i} < L_{j}$. It is easy to check that $S\setminus \{(l_{i},r_{i}),(l_{j},r_{j})\} \cup \{(L_{i},R_{j}),(R_{i},L_{j})\}$ is a structure compatible with $w$ with the same number of base pairs as $S$, a contradiction with the assumption that $w$ is a design for $S$.

To show that the degree condition is also sufficient, we need further definitions and claims.
First, we say that a sequence $w\in\Sigma^{*}$ is \Def{saturable} if there is a saturated structure compatible with $w$. Note that the concatenation of two saturable sequences is also saturable. Then the following claim characterizes the cases when a saturable sequence can be split into saturable sequences.

\begin{clm}\label{l:saturated}
Let $w = \wP \wS$ be a saturable sequence of length $k$. If $\wP$ is saturable, then so is $\wS$.
\end{clm}
\begin{dproof}
Consider a saturated structure $S$ compatible with sequence $w$ and saturated structure $\Sp$ compatible with $\wP$. We will construct a saturated structure $\Ss$ compatible with $\wS$.

Consider a graph $G$ with vertex set $\{1,\dots,k\}$ and edge set defined by pairs in $S\cup \Sp$. Obviously, this graph is a collection of alternating paths (alternating between pairs from $S$ and from $\Sp$, starting and ending with positions in $\wS$) and alternating cyclic paths, and it has a planar embedding such that all vertices lie on a line in their order: pairs in $S$ are drawn as non-crossing arcs above the line and pairs in $\Sp$ as non-crossing arcs below the line. Note that every position in $\wS$ is an end-point of exactly one path in the collection.

Define set of base pairs $\Ss$ by pairing the end-points of the paths in $G$, cf. Figure~\ref{fig:saturated}. We will show that $\Ss$ is a structure. Consider a graph $G'$ constructed by adding pairs in $\Ss$ to $G$. This graph is a collection of cyclic paths. Consider an embedding of $G'$ into plane that extends the planar embedding of $G$ by adding arcs corresponding to the pairs in $\Ss$ below the line containing all the vertices. If two base pairs $b,b'\in \Ss$ cross then the cyclic path containing $b$ and the cyclic path containing $b'$ intersect in exactly one point. By Jordan's curve theorem, this is a contradiction. It follows that $\Ss$ is a saturated structure, and hence $\wS$ is also saturable.
\end{dproof}

We define $w$ to be an \Def{atomic saturable sequence} if no proper prefix
of $w$ is saturable. Clearly, every saturated structure compatible with an atomic saturable sequence $w$ contains the base pair $(1,|w|)$. On the other hand, by Claim~\ref{l:saturated}, if every saturated structure compatible with $w$ contains the pair $(1,|w|)$, then $w$ is an atomic saturable sequence. A design $w$ that is also an atomic saturable sequence will be called an \Def{atomic saturable design}.
\begin{figure}[tb]
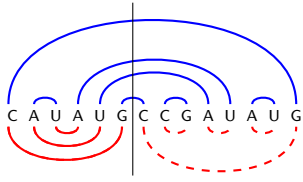

\centering
\begin{minipage}{0.40\textwidth}
\centering
  {\JRNA[6/7/.8/1.5]{CAUAUGCCGAUAUG}{1/14,2/3,4/11,5/10,6/7,8/9,12/13}{1/6,2/5,3/4}{1/6,2/5,3/4,7/14,10/11,8/9,12/13}{}}
\end{minipage}
\begin{minipage}{0.55\textwidth}
\caption{Construction the saturated structure compatible with the suffix $\wS$. The vertical line splits the sequence into a prefix $\wP$ and a suffix $\wS$. Blue and red arcs depict saturated structures compatible with $w$ and $\wP$ respectively. Dashed red arcs represent the induced saturated structure compatible with $\wS$.}
\label{fig:saturated}
\end{minipage}
\end{figure}
A concatenation of two or more atomic saturable designs is obviously not an atomic saturable sequence and it is not necessarily a design. However, we have the following claims.
\begin{clm}\label{l:atomic1}
The concatenation of $t$ atomic saturable designs $w^{1} \ldots w^{t}$ for structures $S^1 \ldots S^{|t|}$, such that $w^{i}_{1}\ne w^{j}_{1}, \forall 1\le i < j\le t$, is a design for the concatenated (saturated) structure $S=S^1\cdots S^{|t|}$.
\end{clm}
\begin{dproof}
Assume that $W := w^{1}\cdots w^{t}$ is not a design, then there exist a saturated structure $S'\neq S$ for $W$. We show that positing such an alternative structure leads to a contradiction, reminding that each $S^{i}$ is saturated and contains a pair $(1, |w^{i}|)$. 
Assume that there exists a leftmost word $w_i$, $i\in[1,|t|]$, such that $w^{i}_1$ is not paired with $w^{i}_{|w^{i}|}$ in $S'$. 
If $w^{i}_1$ is not paired, then $S'$ is not saturated, a contradiction. Let $w^{j}_{k}$, $j\ge i$, be the partner of $w^{i}_1$ in $S'$, and let $u := w^{i}\cdots w^{j-1}\Prefix{k}{w^{j}}$. If $k = |w^{j}|$, then $j > i$ and, by complementarity, $w^{i}_{1} = w^{j}_{1}$ which contradicts the preconditions. Hence, we can assume that $k < |w^{j}|$. Since $u$ and each of the $w^{i},\dots,w^{j - 1}$ are saturable, by iterated application of Claim~\ref{l:saturated}, we conclude that $v = \Prefix{k}{w^{j}}$ is saturable as well and, from Claim~\ref{l:saturated}, so is $v' = \Substring{k+1}{|w^{j}|}{w^{j}}$. This contradicts the hypothesis that $w^{j}=v.v'$ is an atomic saturable design, since there exists a saturated folding for $w^{j}$ which does not pair its extremities. Consequently, $S'$ pairs the first and last letters in each $w^{k}$, hence $S' = S$ since each $w^{i}$ is a design, again a contradiction. 
We conclude that no alternative saturated folding exists for $W$, i.e. $W$ is a design for $S$.
\end{dproof}
\begin{clm}\label{l:atomic2}
  Consider $t$ atomic saturable designs $w^{1}=w^{1}_1\cdots w^{1}_{|w^{1}|}$, \dots, $w^{t}=w^{t}_1\cdots w^{t}_{|w^{t}|}$ and a pair $a,b$ of complementary letters such that $w^{i}_{1}\ne b$ for every $1\le i\le t$ and $w^{i}_{1}\ne w^{j}_{1}$ for every $1\le i < j\le t$. Then $W = aw^{1}\cdots w^{t}b$ is an atomic saturable design.
\end{clm}
\begin{dproof}
We will first show that $W$ is an atomic saturable sequence. Assume to the contrary that there is a proper prefix of $W$ that is saturable. Consider the shortest such prefix $aw^{1}\cdots w^{i}\Prefix{j}{w^{i + 1}}$. Obviously, $a$ has to be paired with $w_{j}^{i + 1}$, otherwise we can find a shorter saturable prefix. This implies that $b = w_{j}^{i + 1}$ and that $w^{1}\cdots w^{i}\Prefix{j - 1}{w^{i + 1}}$ is saturable as well. By repeated application of Claim~\ref{l:saturated}, we have that $\Prefix{j - 1}{w^{i + 1}}$ is saturable. Since it is a prefix of atomic saturable sequence $w^{i + 1}$, it must be the empty sequence, i.e., $j = 1$. Therefore, $b = w_{1}^{i + 1}$, a contradiction with the assumptions of the claim. Thus, $W$ is an atomic saturable sequence.

Now we will show that $W$ is a design. Consider any MFE (saturated) structure $S$ for $W$. Since $W$ is atomic saturable, $a$ is paired with $b$ in $S$. By Claim~\ref{l:atomic1}, $w^{1}\cdots w^{t}$ is a design. It follows that $W$ is a design as well.
\end{dproof}

\label{th:saturated}To prove the sufficiency of the degree condition, consider the following algorithm, which takes as input a saturated structure $S$ with $\MaxDeg{S}\le 2c$, and returns a design $w$ for $S$:
\begin{itemize}
\item 
  Let $\{[l_{i} ,r_{i}]\}_{i = 1 }^{d}$ be the children of the root. Assign to each $w_{l_{i}},w_{r_{i}}$ complementary bases such that $\forall 1\le i < j\le d:\ w_{l_{i}}\neq w_{l_{j}}$.
\item 
  While there exists an unprocessed internal node $[a,b]$ whose parent has been processed  (if there is no such node, stop and return $w$). Let $\{[l_{i} ,r_{i}]\}_{i = 1 }^{d}$ be the children of $[a,b]$. Assign to each $w_{l_{i}},w_{r_{i}}$ complementary bases such that $\forall 1\le i \le d:\ w_{l_{i}}\neq w_{a}$ and $\forall 1\le i < j\le d:\ w_{l_{i}}\neq w_{l_{j}}$.
\end{itemize}

Note that since the alphabet contains $c$ pairs of complementary bases, the assignment at each step of the algorithm is possible. We will show that the returned sequence $w$ is a design for $S$. We will show by tree induction on the size subtrees that $w_{i}\cdots w_{j}$ is an atomic saturable design for every internal node $[i,j]$. It is easy to check that this is satisfied at the leaves. Consider an internal node $u$. By the induction hypothesis, sequences for each child subtree of $u$ are atomic saturable designs. Furthermore, by the choice of bases at children nodes of $u$, all assumptions of Claim~\ref{l:atomic2} are satisfied, hence, the sequence for node $u$ is also an atomic saturable design. The claim holds. Finally, we can apply Claim~\ref{l:atomic1} at the root, which yields that $w$ is a design.
\end{dproof}

\begin{cor}[Result \Result{2}]\label{t:two-letter}
  A structure $S$ is $\Sigma_{1,0}$-designable if and only if it does not contain any base pairs, or it is saturated and $\MaxDeg{S}\le 2$.
\end{cor}

\begin{dproof}
  If $S$ contains a base pair and an unpaired position, then it can be easily checked that $S$ is not $\Sigma_{1,0}$-designable.   
Hence, any $\Sigma_{1,0}$-designable structure is either empty, and trivially designable using a single letter, or saturated. In the latter case, by Theorem~\ref{t:result4}, we know that designable structures are exactly those that are saturated, and such that $\MaxDeg{S}\le 2$. The claim follows.  
\end{dproof}

\begin{cor}[Result \Result{3}]\label{t:three-letter}
  A structure $S$ is $\Sigma_{1,1}$-designable if and only if $\MaxDeg{S}\le 2$.
\end{cor}

\begin{dproof}
  First, suppose $S$ is $\Sigma_{1,1}$-designable and let $w$ be a design for $S$. 
Then $\Paired{w,S}$ is a design for $\Paired{S}$. Since the paired restriction $\Paired{S}$ is saturated, it is over alphabet $\Sigma_{1,0} \subset \Sigma_{1,1}$, and by Theorem~\ref{t:result4}, $D(\Paired{S})\le 2$. Hence, $\MaxDeg{S} = \MaxDeg{\Paired{S}}\le 2$.

  Conversely, suppose that $\MaxDeg{S}\le 2$. Construct a design for $S$ as follows. Since $\Paired{S}$ is saturated, by Theorem~\ref{t:result4}, there is a design $\bar w$ for $\Paired{S}$ over $\Sigma_{1,0} \subset \Sigma_{1,1}$. Construct $w$ from $\bar w$ by inserting the base without a complementary base at every unpaired position of $S$. Let $S'$ be an MFE structure for $w$. Obviously, all unpaired positions in $S$ are also unpaired in $S'$. We must have $\Paired{S'} = \Paired{S}$, otherwise we have an alternative structure for $\bar w$, a contradiction. Hence, $S' = S$, i.e., $w$ is a design for $S$.
\end{dproof}

Result \Result{4} follows readily from Theorem~\ref{t:result4} by taking $c = 2$.

\begin{lem}[Result \Result{5}]
  Any structure that contains $\MotifA$ or $\MotifB$ is not $\Sigma_{2,0}$-designable.
\end{lem}

\begin{dproof}
  Assume that $S$ is $\Sigma_{2,0}$-designable and let $w$ be a design for $S$. Then $\Paired{w,S}$ is a design for $\Paired{S}$. Since $\Paired{S}$ is saturated, by Theorem~\ref{t:result4}, $\MaxDeg{S} = \MaxDeg{\Paired{S}}\le 4$, hence, $S$ cannot contain motif $\MotifA $.
Now, assume to the contrary that $S$ contain motif $\MotifB $ appearing at node $[a,b]$ of $T_{S}$. Let $\{[l_{i},r_{i}]\}_{i = 1 }^{3}$ be some paired children of $[a,b]$ and the node $[a,b]$ if $[a,b]$ is not the root, and $[u,u]$ an unpaired child of $[a,b]$. Let $L_{i} = l_{i}$ and $R_{i} = r_{i}$ if $[l_{i},r_{i}]$ is a child of $[a,b]$, and $L_{i} = r_{i}$ and $R_{i} = l_{i}$ if it is $[a,b]$. If among bases $w_{L_{1}},\dots,w_{L_{3}}$ there is a pair of repeated letters, then we can construct an alternative MFE structure for $w$ (see the first paragraph in the proof of Theorem~\ref{t:result4}). Assume that these three bases are different. Then for some $i = 1,2,3$, $w_{u}$ equals either $w_{l_{i}}$ or $w_{r_{i}}$, say it equals $w_{l_{i}}$. Then $S\setminus \{(l_{i},r_{i})\} \cup \{(u,r_{i})\} $ is an MFE structure for $S$, a contradiction with the assumption that $w$ is a design for $S$.
%
\end{dproof}

\begin{thm}[Result \Result{6}]
  \label{t:result6}
  If the tree representation of a structure $S$ admits a separated coloring then $S$ is $\Sigma_{2,0}$-designable.
\end{thm}
\begin{dproof}
Given a sequence $w$, we define the \Def{level} $\Level{i}$ of position $i$ as $\Level{i} = |\Prefix{i}{w}|_{\Gb}-|\Prefix{i}{w}|_{\Cb}$. 

\begin{clm}
  \label{lem:equalheight}
  Consider any structure compatible with sequence $w$ that contains some $\Ab-\Ub$ base pair between positions at different levels, then some $\Gb$ or $\Cb$ is left unpaired.
\end{clm}

\begin{dproof}
  Consider that the $\Ab-\Ub$ base pair occurs at position $(a,b)$, and note that the bases of the substring $\Substring{a+1}{b-1}{w}$ can only base pair among themselves without introducing crossings. We will show that $\Gb$'s and $\Cb$'s are not balanced in this substring. Since $w_{b}\in\{\Ab,\Ub\}$, $\Level{b}=\Level{b-1}$. Hence, by the definition of $\LevelFun$, we have that 
  $$|\Substring{a+1}{b-1}{w}|_{\Gb}-|\Substring{a+1}{b-1}{w}|_{\Cb}= \Level{b-1}-\Level{a} = \Level{b}-\Level{a}\neq 0\,.$$
  Therefore, at least one $\Gb $ or $\Cb $ in the substring remains unpaired in this structure.
\end{dproof}

  Consider a separated coloring of the tree representation of $S$. We will use this coloring to construct a design $w$ for $S$, by specifying a nucleotide at each position of $w$. First, for each unpaired position $i$, set $w_{i} = \Ub $. 
Second, apply a modified version of the algorithm described in Theorem~\ref{th:saturated} to set the bases of paired positions in which black nodes are assigned to base pair $\Gb - \Cb $, white nodes to $\Cb - \Gb $ and grey nodes to $\Ab - \Ub $ or $\Ub - \Ab $. The algorithm ignores unpaired nodes in the tree representation of $S$. Since the coloring is proper such assignment is always possible at every step of the algorithm. We claim that for any node $[i,j]$ (paired or unpaired), the level of position $i$ is the same as the level of the node $[i,j]$. To verify this, observe that  the substring of $w$ corresponding to any subtree has the same number of $\Gb $'s and $\Cb $'s. Hence, for any node $[i,j]$, the level of position $i$ depends only on nodes on the path from this node to the root. It is easy to check that the level of $i$ is equal to the level the node. Note that if $[i,j]$ is a grey node then the level of position $j$ is the same as the level of $i$, i.e., the same as the level of $[i,j]$.

We will show that the constructed $w$ is a design for $S$. Since all $\Cb $'s and  $\Ab $'s of $w$ are paired in $S$, $S$ is an MFE structure for $w$. We need to show that it is the only MFE structure for $w$. Consider an MFE structure $S'$ for $w$ different from $S$. Since $w$ has the same number of $\Gb $'s and $\Cb $'s, $S'$ must pair all $\Gb $'s, $\Cb $'s and $\Ab $'s of $w$. We will show that that all unpaired positions in $S$ are also unpaired in $S'$. Assume to the contrary that position $i$ is unpaired in $S$, but it is paired to $j$ in $S'$. We must have $w_{i} = \Ub $ and $w_{j} = \Ab $. Since the coloring is separated, the unpaired node $[i,i]$ has a different level than the grey node containing $j$, and hence, the level of $i$ is different from the level of $j$. It follows by Claim~\ref{lem:equalheight} that some $\Gb $ or $\Cb $ is unpaired in $S'$, a contradiction.
Consider paired restrictions of $S$, $S'$ and $w$.
Both $\Paired{S}$ and $\Paired{S'}$ are saturated and compatible with $\Paired{w,S}$ and they are different since $S$ and $S'$ are different and agree on the unpaired positions. Furthermore, $\Paired{w,S}$ can be produced by the algorithm described in Theorem~\ref{th:saturated} for the input structure $\Paired{S}$, and hence, by Theorem~\ref{t:result4}, $\Paired{w,S}$ is a design for $\Paired{S}$, which contradicts the existence of $\Paired{S'}$. Hence, $w$ is a design for $S$.
\end{dproof}


\begin{thm}[Result  \Result{7}]
  If $w$ is a design for a structure $S$, then for any integer $k\ge 1$, $w^{[k]}$ is a design for $S^{[k]}$. In particular, if a structure $S$ is $\Sigma_{2,0}$-designable, then so is $S^{[k]}$.
\end{thm}

\begin{dproof}
  Consider a designable structure $S$ and let $w = w_{1}\cdots w_{n}$
  be a design for $S$. We will show that
  $w^{[k]}$ is a design for $S^{[k]}$. Let the $i$-th $k$ positions in $S$ be called the
  \emph{region $i$}. Note that the positions in region $i$ of $S^{[k]}$
  correspond to the $i$-th position in $S$.

  First, we will show that $S^{[k]}$ is an MFE structure for $w^{[k]}$.
  Consider an MFE structure $S'$ of $w^{[k]}$. Define an \emph{interaction
    graph} of $S'$, denoted by $I(S') = (V_{I(S')},E_{I(S')})$, as
  follows: the vertex set $V_{I(S')}$ is the set of positions in $w$,
  i.e., $\{1,\dots,n\}$, and there is an edge between $i$ and $j$ in
  $I(S')$ if there exists a pair between a position in region $i$ and a
  position in region $j$ in $S'$. Note that $I(S')$ is a bipartite
  graph: indeed, vertices of any cycle in $I(S')$ are positions in $w$
  that alternate between $\Ab$ and $\Ub$, or between $\Cb$ and
  $\Gb$. Also note that $I(S')$ is an outer-planar graph: base pairs are pairwise non-crossing and can therefore be drawn without crossings on the upper half-plane, leaving the lower half-plane on the outer face. 
  Assign each edge
  $e$ in $E_{I(S')}$ a weight $c(e)$ equal to the number
  of pairs between regions $i$ and $j$ in $S'$. Note that the sum of all
  weights in $I(S')$, denoted as $\|E_{I(S')}\|$, equals $|S'|$. We
  have the following claim.

  \begin{clm}
    \label{cl:matching}
    If $M$ is a maximum matching in $I(S')$ then $|S'|\le
    k|M|$. Moreover, if $|S'| = k|M|$ then every minimum
    vertex cover of $I(S')$ covers every edge exactly once.
  \end{clm}

  \begin{dproof}
    Note that for any vertex $i$ in $V_{I(S')}$, the sum of the
    weights of edges incident with $i$ is at most $k$. Consider a
    smallest vertex cover $C$ of $I(S')$, and take the sum of these
    inequalities over all vertices $i$ in the cover $C$:
    \begin{equation}
      \sum_{i\in C} \sum_{\text{$e$ incident with $i$}} c(e)\le k|C|\,.
      \label{eq:cover-sum}
    \end{equation}
    Since $C$ is a vertex cover, the weight of every edge in
    $E_{I(S')}$ appears at least once on the left side of
    \eqref{eq:cover-sum}, hence $|S'| = \|E_{I(S')}\|\le k|C|$. By K\"onig's
    Theorem, the maximum matching $M$ in $I(S')$ has the same number of
    edges as $C$, i.e., $|S'|\le k|M|$. The equality implies that the
    weight of every edge in $E_{I(S')}$ appears exactly once on the
    left side of \eqref{eq:cover-sum}, i.e., that vertex cover $C$
    covers every edge exactly once.
  \end{dproof}

  Given a matching $M$ in $I(S')$, we can construct a structure
  $S_{M}$ for $w$ with $|M|$ pairs as follows: for every edge
  $\{i,j\}$ in $M$, add pair $(i, j)$. This is a valid
  (pseudoknot-free) structure, since $M$ is a subgraph of outer-planar
  graph $I(S')$. It follows that $|M|\le |S|$. If $M$ is a maximum
  matching on $I(S')$, we have by Claim~\ref{cl:matching} that
    $|S'|\le k|M|\le k|S| = |S^{[k]}|$
  i.e., $S^{[k]}$ is an MFE structure for $w^{[k]}$. It also follows
  that $|S'| = k|M|$ and that $|M| = |S|$. Since $S$ is a unique
  structure for $w$ and $|S_{M}| = |M| = |S|$, we have that $S_{M} =
  S$, i.e., there is only one maximum matching in $I(S')$.
  We need the following claim to show that all connected components in
  $I(S')$ have at most 2 vertices.

  \begin{clm}
    Let $G$ be a connected bipartite graph on at least
    three vertices with unique maximum matching $M$. Then there exists
    a minimum vertex cover of $G$ that covers some edge twice. 
  \end{clm}

  \begin{dproof}
    First, we will show that every vertex in $G$ is incident to an
    edge in matching $M$. Assume the contrary and consider all
    vertices in $G$ which are incident to only non-matching edges. If
    two of these vertices are incident then the matching is not
    maximal. Otherwise, let $u$ be such a vertex and $v$ its
    neighbor. Vertex $v$ must be incident to a matching edge. We can
    construct a new matching by removing this edge and adding edge
    $uv$, which contradicts the assumption that $M$ is a unique maximal
    matching.

    Take a maximal path $P$ alternating between matching and non-matching
    edges in $G$. Let $u$ be an endpoint of $P$ and $e$ the edge on
    $P$ incident to $u$. If $e$ is a non-matching edge then $u$ must
    be incident to a matching edge, say $f$. By maximality of $P$, the
    other endpoint $v$ of $f$ must be on $P$. Since every internal
    vertex of $P$ is incident to a matching on $P$, $v$ must be the
    other endpoint of $P$ and the edge incident to $v$ on $P$ must be
    a non-matching edge. Hence, we have an alternating cycle $P + f$
    which contradicts the uniqueness of the maximal matching. Thus,
    $P$ starts and ends with matching edges. Next, we show that $u$ is
    a pendant vertex (has degree one). Assume to the contrary $u$ is
    incident to another (non-matching) edge $f = uv$. By maximality of
    $P$, $v$ is on $P$, which yields a cycle. If this cycle is even, we have
    an alternating cycle, which contradicts the uniqueness of the
    matching, and if it is odd, we have a contradiction with the fact
    that $G$ is bipartite. Hence, both endpoints of $P$ are pendant.

    Consider a minimum vertex cover $C$ of $G$. By well-known
    K\"onig's theorem, every minimum vertex cover in a bipartite graph
    uses exactly one endpoint of every edge of a maximum matching and
    no other vertices. Since the endpoints of $P$ are pendant, and $G$ is
    connected and has $\ge 3$ vertices, $P$ must have at least
    three edges. Since endpoints of $P$ are pendant and incident
    to matching edges, we can assume that $C$ does not contain
    endpoints of $P$, i.e, contains the second and last by one vertex
    of $P$. It is easy to see that at least one non-matching edge is
    covered twice by $C$.
  \end{dproof}
  Consider a connected component $K$ of $I(S')$. Since $I(S')$ has
  a unique maximum matching, so does $K$. If $K$ has more than two
  vertices, it contains a minimum vertex cover of $K$ that covers some
  edge twice. It follows that there is a minimum vertex cover of
  $I(S')$ that covers some edge twice. Hence, by
  Claim~\ref{cl:matching}, $|S'|\le k|M|$, a contradiction. It follows
  that every connected component of $I(S')$ has at most two vertices,
  hence, either $S'$ is not MFE or $S' = S^{[k]}$.
\end{dproof}


\begin{thm}[Result \Result{8}]
Each structure $S$ without $\MotifA$ and $\MotifB$ can be transformed into a $\Sigma_{2,0}$-designable structure $S'$  by inflating a subset of its base pairs (at most one per band). Furthermore, this transformation can be done in $\Theta(n)$ time.
\end{thm}

\begin{dproof}
We start with the greedy coloring of $T_{S}$. Since $S$ does not contain $\MotifA$ and $\MotifB$, it is a proper coloring and there is no node having both a grey child and an unpaired child. We will insert base pairs within $S$ so that the grey nodes and any unpaired node end up at levels of different parities. If the root has a grey child, assign even parity to the grey nodes, otherwise (if the root has an unpaired child, or no grey and no unpaired children), assign even parity to the unpaired nodes.

Now we proceed from the children of the root towards leaves adjusting parity level for grey and unpaired nodes to keep one type even and the other one odd. We repeatedly apply the following simple operation on  $T_{S}$:
If the node $N$ does not match its intended parity level. Denote $N_P$ the parent of $N$ ($N_P$ is not the root as all children of the root already have the correct parity level) and $N_{PP}$ the parent of $N_P$. Insert a new paired node $N_N$ between $N_{PP}$ and $N_P$, assign it with the color of $N_P$, and apply the greedy algorithm on $N_N$. Observe that $N_P$ always takes either black or white color changing the parity level of all its descendants (including $N$). Note that the children of $N_P$ may get recolored, we can even get one more grey child but after this operation the parity levels of all children of $N$ are correct and we do not change parity levels outside the subtree rooted at $N$. After fixing all nodes, we get a separated proper coloring (which is actually the greedy coloring) of $T_{S'}$. Hence, by Theorem~\ref{t:result6}, $S'$ is designable. Figure~\ref{fig:inflation} illustrates this process.
\end{dproof}

\section{Conclusion, discussion and perspectives}
\label{sec:conclusion}

In this work, we introduced the {\em Combinatorial RNA Design problem}, a {\em minimal} instance of the RNA design problem which aims at finding a sequence that admits the target structure as its unique base pair maximizing structure. First, we provided complete characterizations for the structures that can be designed using restricted alphabets. Then we considered the RNA design under a four-letter alphabet, and provide a complete characterization of designable saturated structures, i.e., free of unpaired positions. Turning to those target structures that contain unpaired positions, we provided partial characterizations for classes of designable/undesignable structures, and showed that the set of designable structures is closed under the stutter operation. Finally, we introduced structure-approximating version of the problem and, assuming that the input structure avoids two motifs, provided a structure approximating algorithm of ratio $2$ for general structures.

An important question that is left open by this work is the computational complexity of the RNA design problem. Schnall-Levin~\emph{et al.}~\cite{Schnall-Levin2008} established the {\sf NP}-hardness of a more general problem, called the inverse Viterbi algorithm, which takes as input a stochastic grammar (representing the energy model) and a targeted parse tree (representing the structure), and outputs a sequence (design) whose most probable parsing should match the target. However this result does not settle the complexity of the RNA design, essentially because the proposed reduction relies critically on an encoding of 3-SAT instances within the input grammar. While the hypothetical {\em perfect} grammar/energy model for RNA folding probably differs from the currently accepted Turner model, it should ultimately reflect the laws of physics and should certainly not depend on the instance. As the reduction~\cite{Schnall-Levin2008} requires a different grammar (i.e., energy model) for each instance, it does not seem easily adaptable into a proof that holds for a fixed energy model. Consequently, despite two decades of work on the subject, the computational tractability of RNA design is still open, either in its general instance and in our combinatorial version.

%
%
%
Besides complexity issues, natural extensions of this work may include the consideration of more general base pairing models, more realistic energy  models (ideally, the Turner energy model~\cite{Turner2010}), or the design under other objectives, such as the Boltzmann probability~\cite{Zadeh2011}.
However, even the simplest of modifications, allowing $\Gb-\Ub$ base pairs, would invalidate parity properties that are critical to the proofs of some of our results and algorithms. More precise bounds for the ratio of the structure-approximating could be established. Finally, better algorithms could be designed for the problem, attempting to minimize the number of modifications so that a given structure becomes designable (or, more modestly, belongs to an identified class of designable structures).

\bibliographystyle{abbrv}
\bibliography{biblio}{}

\ShowTODO{
\newpage
\appendix
\setcounter{tocdepth}{5}


\listoftodos}


\end{document}